\documentclass[sigconf]{acmart}
\usepackage{siunitx}
\usepackage{algorithm}
\usepackage{algpseudocode}
\usepackage{multirow}
\usepackage{subfig}
\copyrightyear{2027}
\acmYear{2027}
\setcopyright{cc}
\setcctype{by}
\acmConference[ASPDAC '27]{32nd Asia and South Pacific Design Automation Conference}{January 25--28, 2027}{Tokyo, Japan}
\acmBooktitle{32nd Asia and South Pacific Design Automation Conference (ASPDAC '27), January 25--28, 2027, Tokyo, Japan}
\acmDOI{10.1145/3840404.3848378}
\acmISBN{979-8-4007-2923-2/2027/01}

\AtBeginDocument{%
  }

\begin{document}

\title{
qLevelPlace: A Level-Based Placement Framework for RSFQ Circuits
}

\author{Ziyu Liu}
\email{zliu4130@usc.edu}
\affiliation{%
  \institution{University of Southern California}
  \city{Los Angeles}
  \state{California}
  \country{USA}
}

\author{Jingkai Hong}
\email{jingkaih@usc.edu}
\affiliation{%
  \institution{University of Southern California}
  \city{Los Angeles}
  \state{California}
  \country{USA}
}

\author{Robert S. Aviles}
\email{rsaviles@usc.edu}
\affiliation{%
  \institution{University of Southern California}
  \city{Los Angeles}
  \state{California}
  \country{USA}
}

\author{Sasan Razmkhah}
\email{razmkhah@usc.edu}
\affiliation{%
  \institution{University of Southern California}
  \city{Los Angeles}
  \state{California}
  \country{USA}
}

\author{Peter A. Beerel}
\email{pabeerel@usc.edu}
\affiliation{%
  \institution{University of Southern California}
  \city{Los Angeles}
  \state{California}
  \country{USA}
}

\author{Massoud Pedram}
\email{pedram@usc.edu}
\affiliation{%
  \institution{University of Southern California}
  \city{Los Angeles}
  \state{California}
  \country{USA}
}

\renewcommand{\shortauthors}{Liu et al.}

\begin{abstract}
Rapid Single Flux Quantum (RSFQ) logic can operate at ultra-high clock frequencies; however, its physical design is complicated by explicit splitter trees and deep gate-level pipelines in which every logic gate is clocked.
To address this challenge, we present qLevelPlace, a level-based placement framework for RSFQ circuits. We separate gate and splitter placement into two stages to better support level-guided gate-to-row assignment and use a timing-aware quadratic cost function to penalize long interconnects. Experiments on the benchmark circuits show that, on average, qLevelPlace achieves a 2.65$\times$ higher clock frequency and a 51\% reduction in layout area compared with the state-of-the-art RSFQ placer qPlace.
\end{abstract}

\maketitle

\section{INTRODUCTION}

Rapid single-flux-quantum (RSFQ) circuits provide ultra-low energy dissipation on the order of \SI{2e-19}{\joule\per\bit}\cite{holmes2013energy} and have demonstrated operating frequencies of 25--50 GHz~\cite{4277821,7467490,7792366,10531089}. These characteristics position RSFQ as a promising post-CMOS technology for high-performance computing, particularly in domains such as quantum system control \cite{jokar2022digiq}, neuromorphic computing \cite{ishida2020supernpu}, and high-performance computing on a datacenter-scale \cite{holmes2013energy, choi2024supercore}. However, RSFQ design poses unique challenges because every logic gate is clocked and fanout requires explicit non-clocked splitter gates.

Addressing these design challenges has led to the development of several RSFQ clocking schemes, including the widely used single-phase approach and alternatives such as concurrent-flow~\cite{11196919} and multi-phase clocking~\cite{9912068,aviles2024efficient}. In this work, we focus on single-phase clocking, where interconnect delay—particularly from long gate-to-gate connections—often dominates the clock period, making placement a timing-critical stage of the design flow.

Existing RSFQ placement algorithms fall into two main categories: level-based methods and CMOS-style analytical placers. Level-based methods~\cite{9381512,9162467,10546887,9380440,10081391,11196919} exploit the logical pipeline structure of RSFQ circuits by mapping each logic level to exactly one physical row. However, these approaches were developed for concurrent-flow clocking, in which timing is enforced locally at each logic stage. As a result, they neither target single-phase clocking nor control the longest gate-to-gate interconnect. Moreover, the one-row-per-level constraint leads to poor area utilization for highly unbalanced levels. 

\begin{figure}[thbp]
    \centering
    \vspace{-2em}
    \subfloat[Comparison of gate-to-gate interconnect length distributions on benchmark circuit C5315\label{fig:wirelength-distribution}]{%
    \includegraphics[width=0.82\columnwidth]{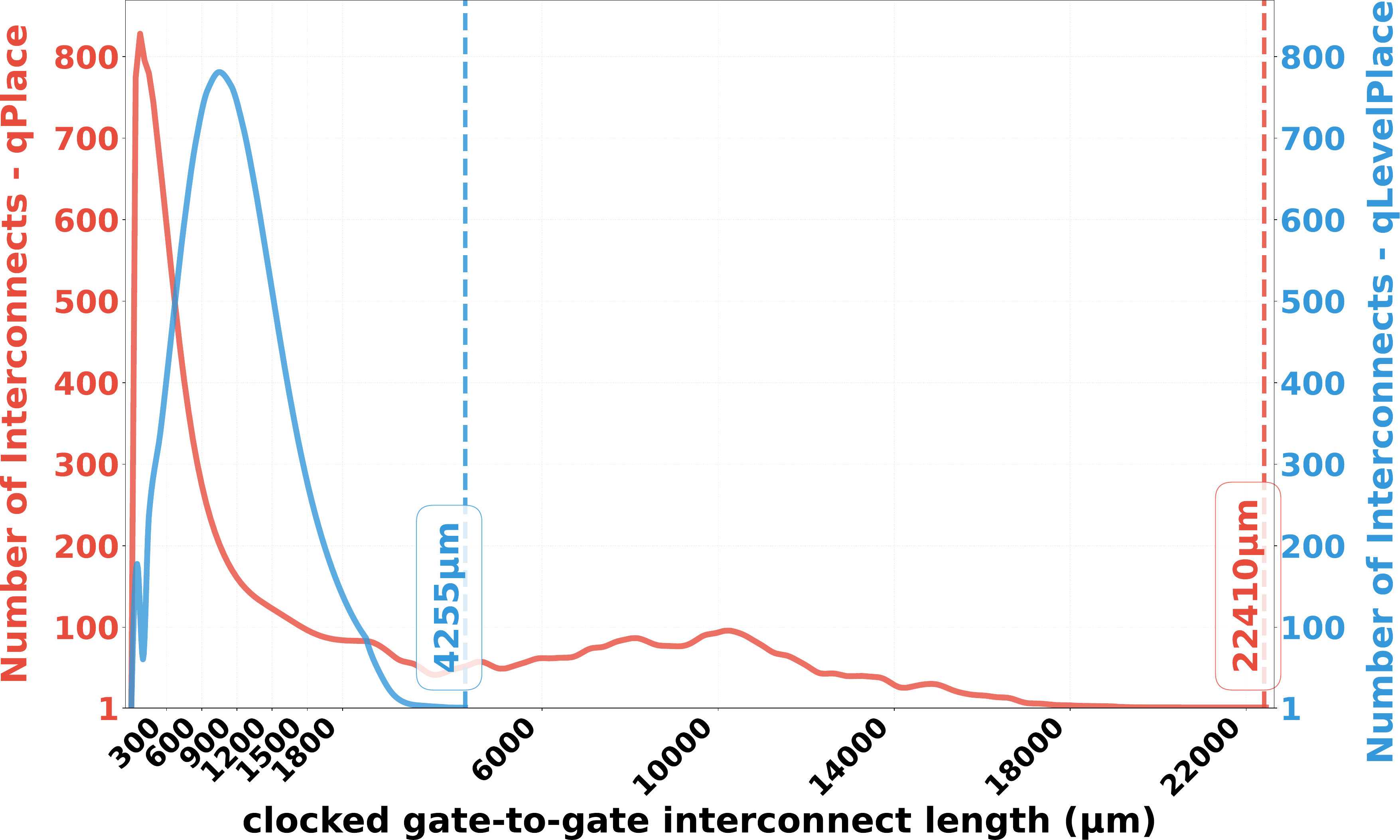}%
    }\\
    \subfloat[CMOS circuit with a single timing path (shown in red)\label{fig:cmos-example}]{%
    \includegraphics[width=0.75\columnwidth]{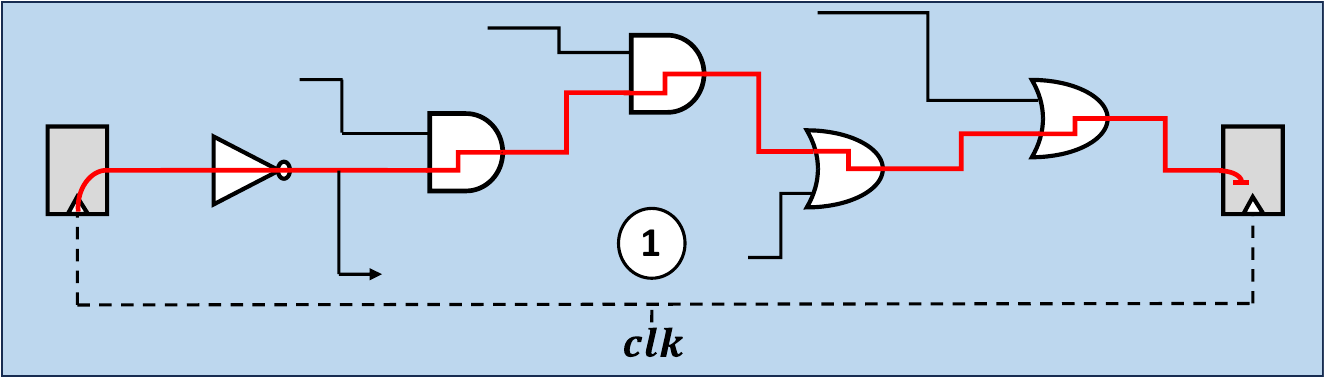}%
    }\\
    \subfloat[RSFQ counterpart with six timing paths (shown in red)\label{fig:rsfq-example}]{%
    \includegraphics[width=0.75\columnwidth]{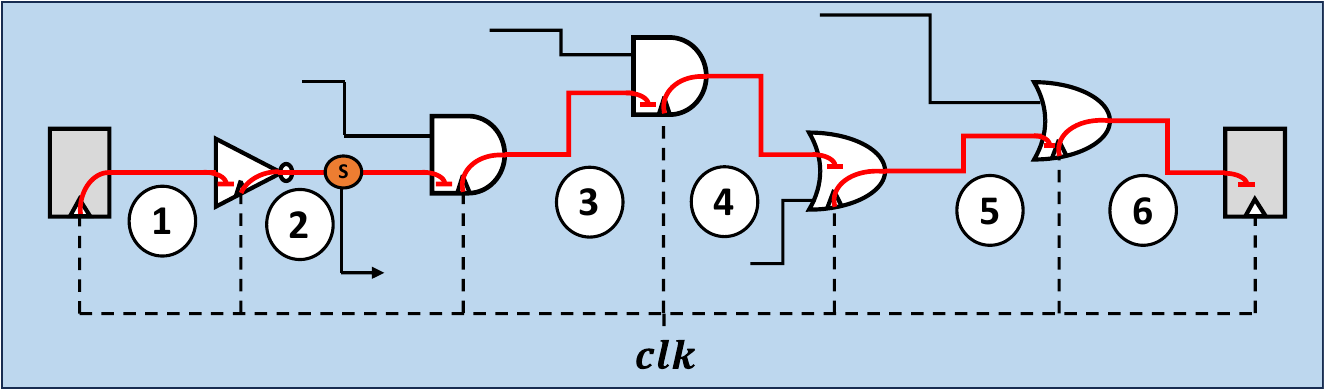}%
    }\\
    \caption{Motivation for RSFQ-specific placement}
    \Description{Three subfigures showing (a) wire length distribution comparison, (b) CMOS timing path example, and (c) RSFQ timing path example demonstrating the motivation for specialized placement algorithms.}
    \label{fig:motivation}
\end{figure}

The state-of-the-art single-phase RSFQ placer qPlace~\cite{7867080} adopts a CMOS-style analytical placement algorithm\allowbreak{}\cite{6106735}
that minimizes the half-perimeter wirelength (HPWL) via smooth approximations of the L$_1$ norm. However, as illustrated in Fig.~\ref{fig:wirelength-distribution}, this 
approach produces distributions with long tails (red curve), indicating numerous outlier long wires. Although such outliers are tolerable in CMOS, where delay accumulates over many gates (Fig.~\ref{fig:cmos-example}), they are critical in RSFQ's gate-level pipelined structure (Fig.~\ref{fig:rsfq-example}), in which each gate-to-gate connection defines an independent timing path, and a single long wire can limit the achievable clock frequency.

In this paper, we propose qLevelPlace, a level-based RSFQ placement framework that addresses the challenges of placing fully path-balanced (FPB) RSFQ circuits using single-phase clocking. Our approach focuses on performance optimization while considering area utilization and routability under limited routing resources. As illustrated in Fig.~\ref{fig:wirelength-distribution}, our method (blue) yields a tighter and more uniform wire-length distribution with fewer long critical connections, enabling higher frequency.
{Our main contributions are:}
\begin{itemize}
    \item \textbf{Two-stage placement} that separates logic gate placement from splitter placement, enabling critical gate-to-gate interconnect length (CIL) minimization.
    \item \textbf{Flexible level-to-row assignment} that allows a logic level to span multiple physical rows, improving area utilization over rigid one-level-to-one-row assignment schemes.
    \item \textbf{Timing-aware edge weighting} that derives edge weights from normalized gate-to-gate delays to prioritize critical connections, improving circuit frequency while reducing estimated CIL.
\end{itemize}
In experiments, qLevelPlace improves clock frequency by up to $3.81\times$ (2.65$\times$ on average) over qPlace~\cite{7867080}. Sections 2--5
respectively review RSFQ placement, present qLevelPlace, evaluate it,
and conclude the paper.

\section{BACKGROUND AND RELATED WORK}
\label{sec:BACKGROUND-AND-RELATED-WORK}

\subsection{RSFQ Logic}
\label{subsec:RSFQ-logic}

RSFQ logic gates operate on quantized magnetic flux, where information is encoded as single flux quantum (SFQ) pulses. Each pulse represents a binary "1", while the absence of a pulse represents a binary "0". From the perspective of physical design, RSFQ circuits exhibit several distinctive characteristics stemming from this pulse-driven nature:

\textbf{Limited fanout capability:} 
Unlike CMOS, where one gate can drive multiple sinks via a hyper-net, 
RSFQ gates can drive a single load. When fanout $>1$, asynchronous splitter gates must be inserted to duplicate the SFQ pulse.
Synthesis tools typically limit this fanout to improve performance \cite{10224066,Aviles2025_ICCAD}, and our placement framework respects this constraint by assuming a maximum fanout of four per gate.

\textbf{Path balancing requirements:} 
Under single-phase synchronous clocking, every gate is clocked and all logic paths must be path-balanced.
The logic level $l(g)$ of gate $g$ is recursively defined as $l(g) = \max_{f \in \text{fanin}(g)} l(f) + 1$, with $l(g)=0$ for primary inputs. FPB netlists ensure all input paths to any gate traverse the same number of logic gates by inserting path-balancing clocked buffers during synthesis.\footnote{Path-balancing clocked buffers are treated the same as regular logic gates during placement.}
Importantly, the throughput of an FPB circuit is inversely proportional to its clock period, 
which is set by the critical-path delay of a single logical stage, comprising a driver gate's clock-to-Q delay, a receiver gate's setup delay, the interconnect delay between the gates, and up to two splitter delays for a fanout-of-four.
This is in sharp contrast to typical CMOS digital circuits in which the critical path passes through 10\,--\,30 gates and their associated interconnects. This dramatic difference motivates our novel RSFQ placement algorithm, which prioritizes minimizing outlier long wires that could otherwise limit clock frequency.

\textbf{Passive Transmission Lines (PTLs):} \label{para:ptl}
RSFQ interconnects use superconducting striplines with limited routing layers; PTLs require no repeaters, and their propagation delay scales linearly with length at approximately $1\,\mathrm{ps}$ per $100\,\mu\mathrm{m}$,
corresponding to a signal velocity of ${\sim}c/3$~\cite{mukhanov2011energy}. 
bulky inductors and limited fabrication scalability yield large gate-to-gate distances, so despite this high propagation speed, interconnect delay often dominates the clock period.

\subsection{RSFQ Placement Problem Definition}
\label{subsec:problem-formulation}

An RSFQ circuit is represented by a directed graph $G = (V, E)$, where $V = V_{\text{logic}} \cup V_{\text{SPL}} \cup V_{\text{PI}} \cup V_{\text{PO}}$, where $V_{\text{logic}}$ is the set of clocked logic gates, $V_{\text{SPL}}$ is the set of splitters, $V_{\text{PI}}$ the set of primary inputs, and $V_{\text{PO}}$ is the set of primary outputs; $E$ is the set of nets.

The placement problem assigns valid coordinates $(x_i, y_i)$ to
all gates under non-overlap constraints, and aims to optimize the geometric
characteristics that impact critical-path delays.

\subsection{Related Work}
\label{subsec:related-work}
Existing placement approaches for RSFQ fall into two categories:

\textbf{CMOS-style Placement:}
Placement in CMOS has been extensively studied, with approaches broadly classified into stochastic methods, e.g., simulated annealing (SA)~\cite{17235}, and analytical methods based on continuous optimization~\cite{6106735}.  
Some RSFQ placers, such as qPlace \cite{7867080}, adapt analytical placement methods~\cite{6106735} that iteratively solve a reweighted quadratic programming problem 
whose objective function approximates HPWL. More precisely, 
the objective function for iteration $k$ is formulated as:
\begin{align}
\text{minimize} \quad 
& \sum_{(i,j) \in E} \left( 
  c_{ij}^{x(k)} (x_i - x_j)^2 
  + c_{ij}^{y(k)} (y_i - y_j)^2 \right) 
\label{eq:cmos_quad_obj}\\
\text{subject to} \quad 
&v_i, v_j \text{ do not overlap}, 
\quad \forall v_i \neq v_j \in V.
\label{eq:overlap}
\end{align}
where $c_{ij}^{x(k)}$ and $c_{ij}^{y(k)}$ are connection weights iteratively updated by the Bound2Bound (B2B) net model~\cite{6106735} based on the placement of the previous iteration. 
This iteratively reweighted quadratic formulation approximates L$_1$-norm HPWL~\cite{6106735}. 

\textbf{Level-based placement: }
Other RSFQ works~\cite{gaj1997timing,10081391,10546887,11196919} adopt level-based placement methods 
which organize gates into rows\footnote{Some works use the term "columns" to describe alternative vertical arrangements.} only according to their logic level.
Gates at the same logic level, which correspond to the same clocked logic stage, are physically aligned on the same placement row. These methods employ concurrent-flow or wave-pipelined clocking schemes~\cite{gaj1997timing,10081391,10546887,11196919}, which maintain data-path and clock-path synchronization using specialized clock distribution circuits.

Our formulation also exploits the levelized structure of RSFQ circuits and solves the placement problem through a structured quadratic optimization framework.

\begin{figure}[t]
    \centering
    \includegraphics[width=0.8\linewidth]{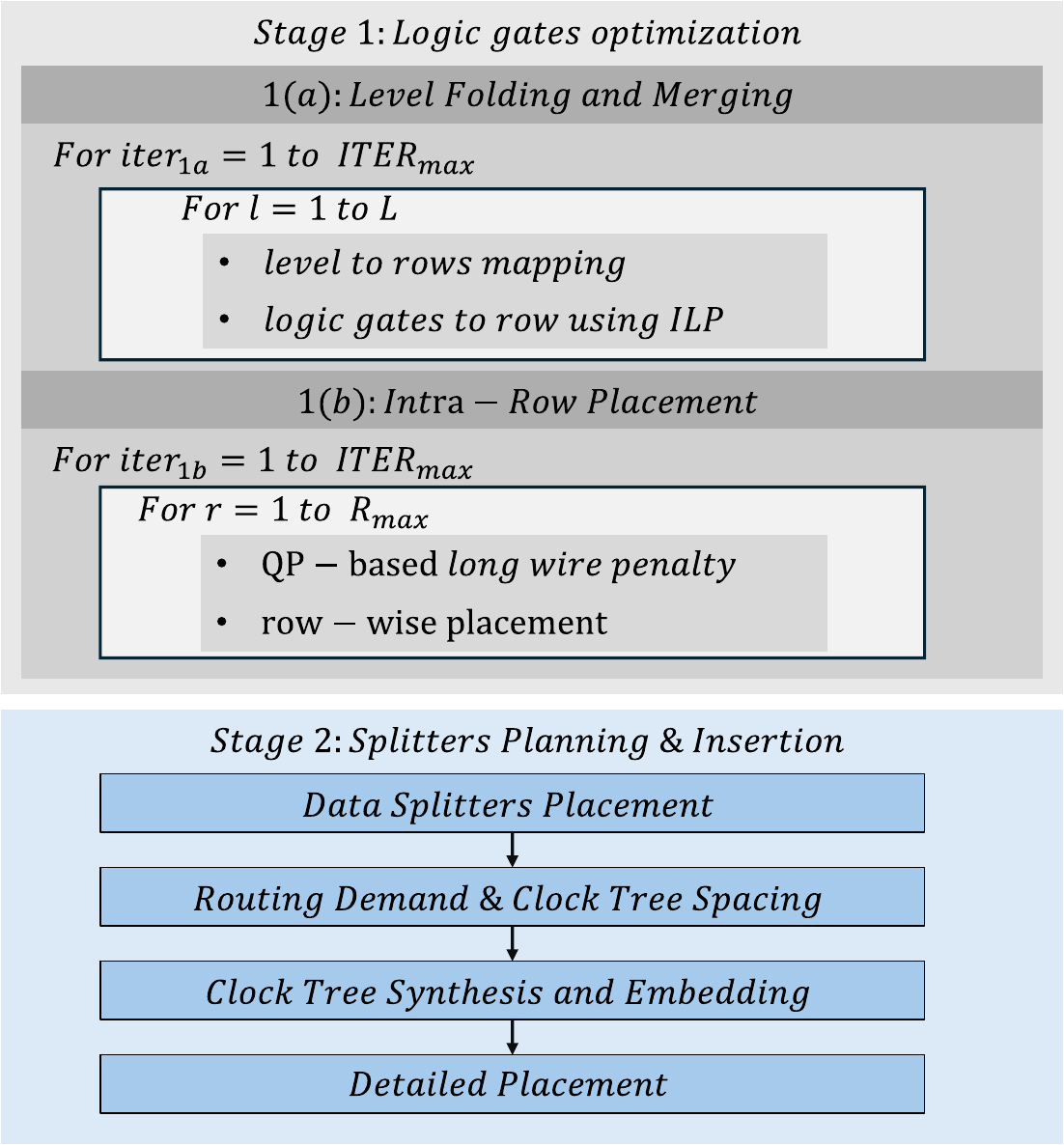}
    \caption{qLevelPlace two-stage algorithm flow}
    \label{fig:qlevel-algo}
    \vspace{-3ex}
\end{figure}

\section{OUR PROPOSED METHOD}
We propose a two-stage placement approach for single-phase RSFQ circuits that decouples the optimization of clocked logic gates from that of unclocked splitters. As illustrated in Fig.~\ref{fig:rsfq-example}, RSFQ's gate-level clocking creates short pipeline stages along each timing path, where stages contain only direct gate-to-gate interconnects or splitter-based fanout.
The placement of logic gates impacts critical timing paths and can benefit from timing-driven placement (Stage 1), while splitters are unclocked fanout elements that can be optimized independently after logic gate positions are fixed (Stage 2). 

\subsection{Overview of Two-Stage Approach}

Recall that all logic gates 
are organized into discrete levels, with each level $i$ exclusively driving level $i+1$. We exploit this inherently levelized structure by dividing the Stage 1 task of logic gate placement into two coordinated subproblems as shown in Figure~\ref{fig:qlevel-algo}. First, \textbf{Y-axis row assignment} determines the physical row for each gate. We model this as a level-guided allocation problem, in which each level is assigned one or more rows that are spaced to avoid vertical overlap between gates. Second, \textbf{X-axis placement} optimizes horizontal locations within each row using a quadratic formulation that strongly penalizes long interconnects~\cite{sigl1991analytical}, followed by a legalization step to remove any remaining horizontal overlaps. This decomposition capitalizes on the levelized nature of RSFQ circuits to reduce optimization complexity while preserving solution quality~\cite{viswanathan2007rql}. It also enables algorithms tailored to each subproblem: the Y-direction assignment can be restricted to a small set of candidate rows per logic level due to the inherently discrete level structure, whereas the X-direction placement retains continuous freedom, allowing efficient gradient-based optimization.

Placement of splitters is handled separately in Stage 2, as illustrated in Figure~\ref{fig:qlevel-algo}. Splitters are removed in Stage 1 and then reinserted directly above the driving gate in the wiring channel between placed rows, minimizing added interconnect delay while respecting row-capacity constraints. This two-stage approach enables placing splitters close to their driving gates rather than treating them the same as logic gates. This is critical for high-frequency RSFQ operation, where individual wire delays can directly impact performance.

\subsection{Stage 1(a): Level Folding and Merging} 
We formulate the assignment of logic gates to physical rows
as an integer linear programming (ILP) problem. Each logic gate \(i\) belongs to
a logic level \(l \in \{1, 2, \ldots, L_{\max}\}\), and the physical rows are
indexed by \(r \in \{1, 2, \ldots, R_{\max}\}\), each with Y-coordinate \(Y(r)\)
and capacity \(C\). For each logic level \(l\), we define an allowed row
interval \([r_l^{\min}, r_l^{\max}] \subseteq \{1, \ldots, R_{\max}\}\), where \(r_l^{\min}\) and \(r_l^{\max}\) denote the minimum and maximum row indices
that may be assigned to gates in level \(l\).
Level folding occurs when gates from a single logic level are distributed across multiple rows ($r_l^{\min} \neq r_l^{\max}$), which happens when the total width of gates at level $l$ exceeds the row capacity $C$.
In this case, the level is folded into multiple rows with $k_l=\lceil W_l/C\rceil$, and an ILP assigns gates to rows while respecting capacity constraints.
Level merging occurs when the row intervals of adjacent levels overlap ($r_l^{\max} \ge r_{l+1}^{\min}$).

The objective of the Y-direction assignment is to minimize the total inter-row
interconnect length for each logic level \(l\), computed by summing the cost of
assigning each gate \(i\) in level \(l\) to a row \(r \in [r_l^{\min}, r_l^{\max}]\). Assuming a uniform inter-row spacing, the objective reduces to minimizing the
sum of row index differences (see Eq.~\ref{eq:row_objective}).\footnote{We also considered a min--max objective 
function to minimize the worst inter-row connection, but this 
significantly increases the ILP complexity.}

We optimize one logic level at a time and perform several passes, sweeping levels \(1\) to \(L_{\max}\)
and reassigning gates within their candidate rows (Algorithm~\ref{alg:y-dir-ilp}). A monolithic global ILP is not required in FPB RSFQ circuits because the levelized structure naturally supports an iterative level-wise approach. This enables a multi-pass ILP that keeps the problem size tractable while still providing strong solution quality in practice. For a given level \(l\), the ILP is formulated as follows:
\begin{align}
\min_{x_{i,r},\, r(i)} \;
& \sum_{i \in N^l}
  \Bigg(
      \sum_{p \in FI(i)} |r(i)-r(p)|
    + \sum_{q \in FO(i)} |r(q)-r(i)|
  \Bigg) \label{eq:row_objective} \\
\text{s.t.} \quad
& \sum\nolimits_{r=r_l^{\min}}^{r_l^{\max}} x_{i,r}=1,
  \quad \forall i\in N^l,
  \label{eq:assignment_constraint} \\
& \sum\nolimits_{i\in N^l} w_i x_{i,r}\le C,
  \quad \forall r\in[r_l^{\min},r_l^{\max}].
  \label{eq:capacity_constraint}
\end{align}
\noindent where $N^l$ represents all the nodes in level $l$, $N^r$ represents all the nodes in row $r$ of the placement; $p$ and $q$ denote the fanin and fanout gates of gate $i$. The binary variable $x_{i,r}\in \{0,1\}$ indicates whether gate $i$ is assigned to row $r$. Constraint~(\ref{eq:assignment_constraint}) ensures that each gate is assigned to exactly one row. The variable $w_i$ is the width of cell $i$, constraint~(\ref{eq:capacity_constraint}) enforces the row capacity limit. Note that during the first iteration of the forward pass in Algorithm~\ref{alg:y-dir-ilp}, the fanout nodes' rows have not been assigned yet; an adjacent row $r_l^{\max} + 1$ is assumed to hold all the fanout nodes.

The number of candidate rows per logic level, \( |r_l^{\max} - r_l^{\min}| + 1 \), is determined by the estimated row capacity \(C\), which is computed from the target aspect ratio.  

A key challenge is avoiding the creation of a very small "residue" row where capacity constraint C forces an additional sparsely populated row to accommodate the total width $W_{tot}$ of a level. The residue is computed as $res = W_{tot} \mathit{mod}\, C$. If this remainder falls below the utilization threshold, $res<\epsilon \cdot C$, we slightly increase the capacity by $C'\gets C+\Delta C$ so that the leftover gates can be absorbed into the preceding rows. We apply this residue check across all levels and iteratively adjust $C$ until no level produces a residue smaller than $\epsilon \cdot C$, thereby improving row utilization and eliminating wasted area from sparsely filled rows. 

\begin{algorithm}[t]
\caption{Level-by-level row assignment with local ILP folding and greedy merging}
\label{alg:y-dir-ilp}
\begin{algorithmic}[1]
\Require Netlist $G(V,E)$; cell width $w(\cdot)$; row capacity $C$
\Ensure Row index $\mathrm{row}(i)$ for each gate $i$
\State $\mathrm{row}(i)\gets -1$ for all $i$
\For{$it = 1$ to $ITER\_MAX$}
    \State $r_{\text{cur}} \gets 0$
    \For{$l = 1$ to $L_{\max}$}
        \State $N_l \gets \textsc{getAllCellsAtLevel}(l)$
        \State $W_l \gets \textsc{getTotalCellWidth}(N_l, w(\cdot))$
        \State $k_l \gets \lceil W_l / C \rceil$ \Comment{Desired folds}
        \If{$k_l \ge 2$}
            \State $r_l^{\min} \gets r_{\text{cur}} + 1$, 
            $r_l^{\max} \gets r_{\text{cur}} + k_l$
            \State $x^\star \gets \textsc{ILP}(w(\cdot), C, [r_l^{\min}, r_l^{\max}], \mathrm{row}(\cdot))$
            \State $\mathrm{row}(i) \gets \arg\max_r x^\star_{i,r}, \forall i \in N_l$
        \Else
            \If{$W_{r_{\text{cur}}} + W_l \le C$}
                \Comment{Merge with current row}
                \State $\mathrm{row}(i) \gets r_{\text{cur}}, \forall i \in N_l$
            \Else
                \Comment{Start new row}
                \State $r_{\text{cur}} \gets r_{\text{cur}} + 1$, 
                $\mathrm{row}(i) \gets r_{\text{cur}}, \forall i \in N_l$
            \EndIf
        \EndIf
    \EndFor
\EndFor
\State \Return $\mathrm{row}(\cdot)$
\end{algorithmic}
\end{algorithm}

\begin{algorithm}[!t]
\caption{Intra-Row Placement}
\label{alg:level_placement}
\begin{algorithmic}[1]
\Require Initial placement $\mathcal{P}_0$, netlist $G=(V,E)$, iterations $N_{\text{iter}}$, weight $\lambda$
\Ensure Optimized placement $\mathcal{P}^*$
\State $\text{cost}^* \gets \infty$
\For{$t = 1$ to $N_{\text{iter}}$}
    \For{each row $r \in \mathcal{R}$ with gate set $V_r$}
        \State \Comment{Compute timing-driven edge weights.}
        \State $\overline{D}_r^{(t-1)} \gets \frac{1}{|E_r|}\sum_{(u,v)\in E_r}\widehat{D}_{uv}^{(t-1)}$
\For{each edge $(i,j)\in E_r$}
    \State $\widehat{D}_{ij} \gets t_{c2q,i} + n_{sp}\cdot t_{sp} + t_{\text{setup}} + \alpha \cdot \text{HPWL}_{ij}^{(t-1)}$ \\\Comment{$\alpha$: HPWL-to-delay scaling \hyperref[para:ptl]{(PTL)}}
    \State $c_{ij} \gets \widehat{D}_{ij} / \overline{D_r}^{(t-1)}$
\EndFor 
        \State \Comment{Solve the quadratic objective for row $r$ using CG}
        \State $\Phi_r(x) = 
            (1-\lambda)\!\sum_{(i,j)\in E_r}c_{ij}(x_i-x_j)^2 
            + \lambda\!\sum_{i\in V_r}(x_i-\bar{x}_i)^2$
            \State $x_i^\star \gets \arg\min_{x_i}\Phi_r(x),\quad \forall i \in V_r$
        \State $\mathcal{P} \gets \textsc{Legalize}(V_r,\;x^\star)$
    \EndFor
    \State $\text{avgHPWL} \gets \text{HPWL}(\mathcal{P})/|E|$
    \State $\text{maxIL} \gets \max_{(i,j)\in E} \|\mathcal{P}(i)-\mathcal{P}(j)\|_1$
    \State $\text{cost} \gets \text{avgHPWL} + \text{maxIL}$
    \State \textbf{if} $\text{cost}<\text{cost}^*$ \textbf{then} $\mathcal{P}^*\gets\mathcal{P},\; \text{cost}^*\gets\text{cost}$
\EndFor
\State \Return $\mathcal{P}^*$
\end{algorithmic}
\end{algorithm}

\subsection{Stage 1(b): Intra-Row Placement} 
\hspace*{\parindent}\textbf{Initial Placement Generation:}
With the Y-direction row assignments fixed, an initial intra-row placement is generated as the starting point for subsequent optimization. For each row \(r\), the gates in \(V_r\) are uniformly distributed in a random order and spaced apart by a distance \(s\). This yields an initial placement width
\begin{equation*}
\textstyle
W_{\max} = \max_{r\in\{1,\ldots,R_{max}\}}
\left( \sum_{i\in V_r} w_i + (|V_r|-1)s \right)
\end{equation*}
\noindent where $w_i$ is the width of gate $i$. The total placement height is described in Section~\ref{subsubsec:integrated-routability}.

\textbf{Intra-Row Placement:}
\label{subsec:in-row-placement}
Once an initial intra-row placement is generated, we optimize horizontal positions within each row using an iterative row-by-row process in Algorithm~\ref{alg:level_placement}. The algorithm iterates over all rows, starting with the primary inputs and moving to the primary outputs $N_{\text{iter}}$ times. For each row $r$, only its gates are movable, while connected gates in other rows remain fixed as anchors. For row $r$, we solve the quadratic placement problem:
\begin{align}
\text{minimize} \quad 
& (1-\lambda)\sum_{(i,j) \in E_r} c_{ij}(x_i - x_j)^2 
 + \lambda \sum_{i \in V_r} (x_i - \bar{x}_i)^2
\label{eq:in_row_obj}\\
\text{subject to} \quad 
& y_i = Y_r, \quad \forall i \in V_r
\label{eq:row_fixed}\\
& x_i + w_i \leq x_k, \quad \forall i, k \in V_r : i < k
\label{eq:no_overlap_row}\\
& \bar{x}_i =
\tfrac{1}{2}\bigl(
\max\nolimits_{j\in N_i} x_j +
\min\nolimits_{j\in N_i} x_j
\bigr)
\label{eq:bounding_midpoint}
\end{align}
\noindent
where $V_r = \{i \in V : y_i = Y_r\}$, $E_r = \{(i,j) \in E : i \in V_r \lor j \in V_r\}$, $N_i = \{j \in V : (j,i) \in E \lor (i,j) \in E\}$.
The first term is a quadratic wirelength penalty that reduces total interconnect length while suppressing long gate-to-gate connections. The edge weight $c_{ij}$ is derived from a timing-based delay estimate between cells $i$ and $j$, computed from the previous placement iteration and normalized by the average path delay, where $t_{c2q,i}$, $t_{\mathrm{setup}}$, $n_{sp}$, and $t_{sp}$ denote the clock-to-Q delay of cell $i$, the setup time of cell $j$, the splitter count, and the splitter delay, respectively. The second term anchors each gate toward the midpoint of its neighbors. A larger~$\lambda$ increases the influence of this centering term, while a smaller~$\lambda$ makes the quadratic wirelength penalty dominant.

We first relax Equation~\eqref{eq:no_overlap_row} and solve the quadratic placement objective using conjugate gradient descent~\cite{6106735,viswanathan2007rql}. This produces continuous gate positions that may contain overlaps and serve as target coordinates for subsequent legalization. We then enforce the non-overlap constraint using a Tetris-style legalization algorithm \cite{hill2002method}. Because each optimization pass updates only one row, legalization is performed independently within that row: gates are sorted by their optimal X-coordinates, and placed sequentially from left to right. Each gate is positioned at $\max(x_{\text{optimal}}, x_{\text{bar}})$, where $x_{\text{bar}}$ tracks the right edge of the last placed gate. If the placement exceeds the row boundary, a secondary scan is performed in the opposite direction. This row-by-row optimization keeps all other rows fixed, efficiently reduces overlap while preserving connectivity, and is inspired by the Look-Ahead Legalization (LAL) technique~\cite{kim2012complx,6106735}.

\subsection{Stage 2:  Splitter Placement}
\label{sec:splitter_placement}
\hspace*{\parindent}\textbf{Datapath Splitter Insertion:}
After logic gates are placed, splitters are inserted to satisfy fanout. Prior level-based RSFQ placers~\cite{9381512,9256439,9162467,10081391,11196919} do not specify how data splitters are placed; we introduce a post-placement strategy for inserting splitters.

Given our level-based  structure, we adopt a vertical attachment strategy: splitters are placed directly above their driving gates by extending the layout vertically. This design is motivated by two RSFQ properties: \textit{Timing}: Critical delay is dominated by horizontal gate-to-gate interconnects; the added splitter connection is short and has negligible impact.
\textit{Routability}: Vertical attachment reduces horizontal congestion in inter-level routing channels. 
This approach is effective because our
circuits are designed with shallow fanout trees ($\leq 4$); however, incorporating more sophisticated splitter-tree optimization remains a promising direction for future work.

\textbf{Clock Splitter and Routing Space:}
\label{subsubsec:integrated-routability}
In the MIT-LL SFQ5ee technology~\cite{tolpygo2014fabrication}, while the full process stack contains nine layers, only two metal layers are available for PTL routing, placing tight constraints on available routing resources. Because of these limited routing resources and the need to accommodate clock splitters inserted between rows, efficient use of vertical spacing becomes critical. To address this, we incorporate a congestion-aware model into the Y-axis spacing between consecutive rows. For each net connecting pins $a$ and $b$, we estimate routing demand using pin midpoint locations and calculate the total number of routing tracks required between consecutive rows. The Y-axis spacing between two given rows is then adjusted as:
$S(r_i, r_{i+1}) = S_{base} + \max(\alpha \cdot D_{i,i+1} \cdot P_{w},\, h_{splitter})$
where $D_{i,i+1}$ is the number of nets crossing between consecutive rows, $P_{w}$ is the wire pitch, $h_{splitter}$ is the height of a clock splitter, and $\alpha$ is an empirically-set expansion factor. Here, $S_{base}$ provides the baseline separation needed for data splitters and signal routing, while the $\max$ term reserves additional space for clock splitters and extra routing. This ensures that each channel can accommodate clock distribution structures and signal wiring congestion even in the worst-case scenario.
This approach simplifies the computation of net crossings because each net in an RSFQ circuit connects only two pins and can be routed over cells. Consequently, there is no need for complex Steiner-tree routing structures as in CMOS hyper-nets.

\textbf{Clock Tree Synthesis and Final Legalization:}
After placement converges, we use the same RSFQ clock-tree-synthesis method as in qPlace which is based on a modified DME (Deferred-Merge Embedding) algorithm~\cite{7867080}. Clock splitters are then inserted into the placement and legalized using standard techniques~\cite{ajayi2019openroad,10.1145/3299874.3318012}.

\begin{table}[t]
\centering
\caption{
Comparison with qPlace.
HPWL and CIL are reported in $\mu\mathrm{m}$, area in $\mu\mathrm{m}^2$, and
post-routed frequency in GHz. The average row reports values normalized to
qLevelPlace.
}
\label{tab:qlevel_vs_qplace}
\scriptsize
\setlength{\tabcolsep}{2.2pt}
\renewcommand{\arraystretch}{0.95}
\resizebox{\columnwidth}{!}{%
\begin{tabular}{l r rrrr rrrr}
\toprule
\multirow{2}{*}{\textbf{Ckt.}}
& \multirow{2}{*}{\textbf{\#G}}
& \multicolumn{4}{c}{\textbf{qLevelPlace}}
& \multicolumn{4}{c}{\textbf{qPlace~\cite{7867080}}} \\
\cmidrule(lr){3-6}
\cmidrule(lr){7-10}
&
& \shortstack{\textbf{Tot.}\\\textbf{HPWL}$\downarrow$}
& \textbf{CIL}$\downarrow$
& \textbf{Area}$\downarrow$
& \textbf{Freq.}$\uparrow$
& \shortstack{\textbf{Tot.}\\\textbf{HPWL}$\downarrow$}
& \textbf{CIL}$\downarrow$
& \textbf{Area}$\downarrow$
& \textbf{Freq.}$\uparrow$ \\
\midrule
c1355 & 1,052
& 0.40M & 1.6K & 6.2M  & 29.4
& 1.26M & 6.6K & 21.5M & 12.3 \\

KSA32 & 1,058
& 0.41M & 1.4K & 7.4M  & 25.0
& 1.50M & 6.5K & 15.0M & 11.2 \\

c499 & 1,064
& 0.44M & 1.8K & 7.2M  & 28.0
& 1.30M & 7.5K & 24.0M & 12.6 \\

c880 & 1,486
& 0.73M & 1.4K & 13.0M & 28.5
& 2.00M & 8.0K & 26.4M & 11.9 \\

c432 & 1,550
& 0.79M & 1.7K & 12.0M & 27.0
& 1.20M & 6.6K & 30.0M & 11.5 \\

c3540 & 3,672
& 2.60M & 3.9K & 30.0M & 19.6
& 3.70M & 11.7K & 64.0M & 7.4 \\

c7552 & 6,084
& 5.00M & 3.8K & 85.0M & 18.3
& 18.0M & 22.8K & 103M & N/A \\

c6288 & 6,245
& 4.10M & 2.8K & 57.4M & 20.8
& 5.40M & 12.2K & 106M & 7.1 \\

c5315 & 8,749
& 7.60M & 4.2K & 73.7M & 20.0
& 30.0M & 22.4K & 149M & N/A \\

ALU32 & 13,257
& 17.3M &  5.8K & 164M & 11.8
& 22.4M & 23.8K & 200M & 4.0 \\

Multi32 & 21,204
& 27.6M & 7.9K & 240M & 9.0
& 26.9M & 26.7K & 312M & 3.5 \\

BU32 & 66,962
& 104.4M & 10.6k & 837M & 6.1
& 113.5M & 56.9K & 1,023M & 1.6 \\
\midrule

\textbf{Avg.} & --
& 1.00 & 1.00 & 1.00 & 1.00
& \textbf{2.31}
& \textbf{4.51}
& \textbf{2.03}
& \textbf{2.65} \\
\bottomrule
\end{tabular}%
}
\end{table}

\begin{table}[t]
\centering
\caption{
Ablation study of timing-aware edge weighting. Results are reported as without/with, with normalized ratios (with/without). The top-10 critical stages are selected based on the estimated delays of the placement without weighting and held fixed across both evaluations.
}
\label{tab:timing_weight}
\small
\setlength{\tabcolsep}{3.5pt}
\renewcommand{\arraystretch}{0.95}
\begin{tabular}{lccc}
\toprule
{\footnotesize\textbf{Circuit}}
& \shortstack{\footnotesize\textbf{Tot. HPWL (M)}\footnotesize$\downarrow$}
& \shortstack{\footnotesize\textbf{Avg. CIL (k)}\\\footnotesize\textbf{(Top-10)} $\downarrow$}
& \shortstack{\footnotesize\textbf{Avg. Est. Delay (ps)}\\\footnotesize\textbf{(Top-10)} $\downarrow$} \\
\midrule
c1355  & 0.41 / 0.40 (0.98) & 0.8 / 0.6 (0.80) & 27.4 / 25.8 (0.94) \\
KSA32  & 0.44 / 0.41 (0.93) & 0.7 / 0.7 (0.92) & 31.7 / 31.1 (0.98) \\
c499   & 0.44 / 0.44 (1.00) & 0.8 / 0.7 (0.86) & 27.8 / 26.6 (0.96) \\
c880   & 0.73 / 0.73 (1.00) & 0.9 / 0.7 (0.79) & 31.2 / 29.4 (0.94) \\
c432   & 0.77 / 0.79 (1.03) & 1.1 / 0.9 (0.83) & 32.3 / 30.4 (0.94) \\
c3540  & 2.50 / 2.60 (1.04) & 2.5 / 2.0 (0.82) & 47.4 / 42.9 (0.90) \\
c7552  & 5.60 / 5.00 (0.89) & 3.2 / 2.6 (0.81) & 55.2 / 49.0 (0.89) \\
c6288  & 4.70 / 4.10 (0.87) & 2.3 / 1.9 (0.85) & 43.3 / 39.9 (0.92) \\
c5315  & 7.83 / 7.60 (0.97) & 3.2 / 2.5 (0.77) & 49.1 / 41.8 (0.85) \\
ALU32  & 17.79 / 17.38 (0.98) & 5.4 / 4.4 (0.81) & 71.1 / 60.8 (0.86) \\
Multi32& 27.29 / 27.64 (1.01) & 7.9 / 6.8 (0.86) & 93.4 / 82.1 (0.88) \\
BU32   & 103.0 / 104.4 (1.01) & 11.2 / 8.9 (0.79) & 129.6 / 106.6 (0.82) \\
\midrule
\textbf{Average}
& \textbf{-- / -- (0.97)}
& \textbf{-- / -- (0.83)}
& \textbf{-- / -- (0.91)} \\
\bottomrule
\end{tabular}
\end{table}

\section{EXPERIMENTS AND RESULTS}

\subsection{Experimental Setup and Implementation}

\begin{figure}[t]
    \centering
    \includegraphics[width=0.75\columnwidth]{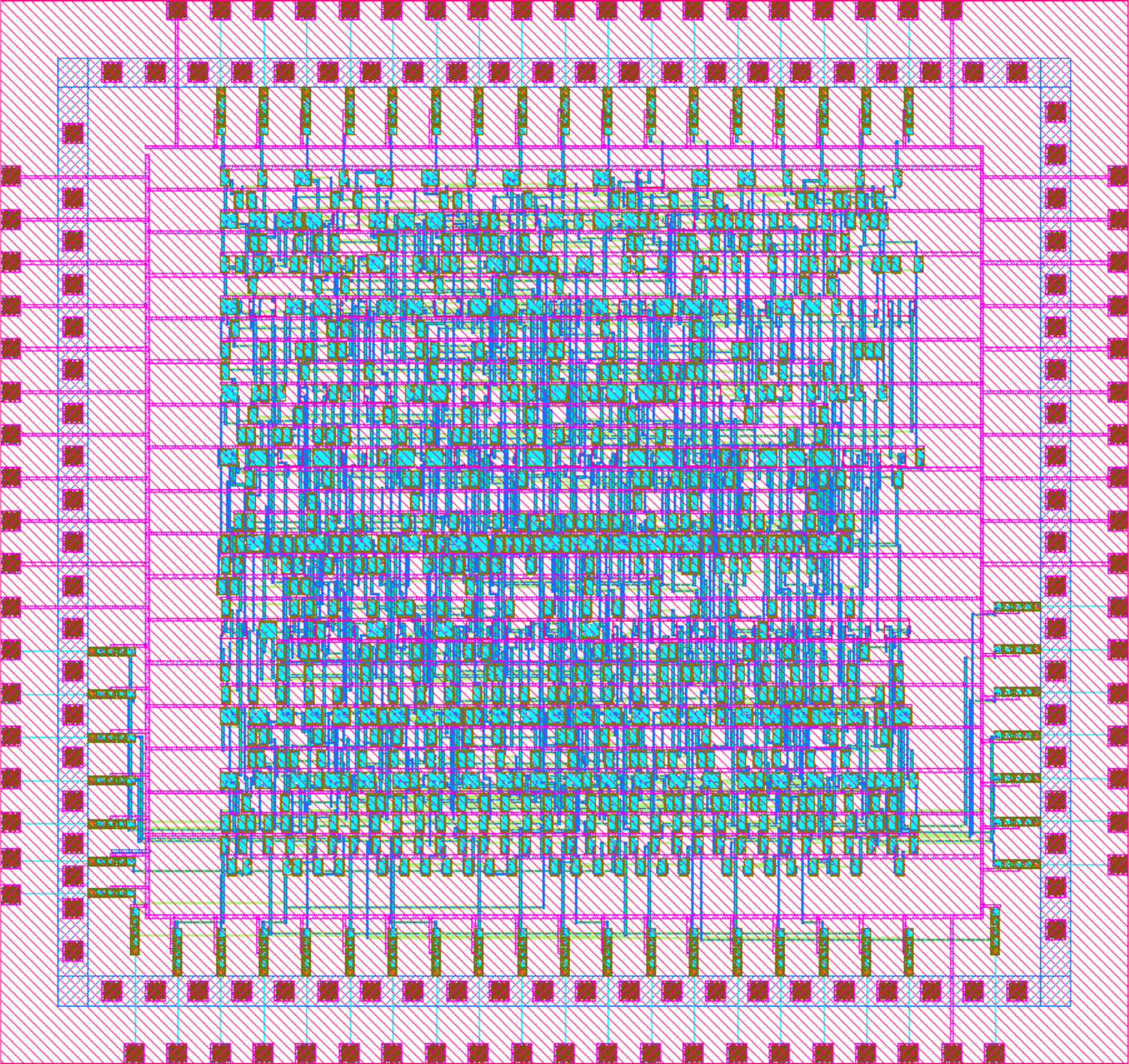}
    \caption{Final GDS layout of benchmark KSA16 based on the placement produced by qLevelPlace.} 
    \label{fig:gds_result}
    \vspace{-1em}
\end{figure}

We evaluated qLevelPlace on the ISCAS-85 benchmark with three additional large circuits: a 32-bit ALU (ALU32), a 32-bit multiplier (Multi32), and a 32-bit butterfly unit (BU32). A sample KSA16 GDS layout is shown in Fig.~\ref{fig:gds_result}. Though small by CMOS standards, the ISCAS-85 circuits are already large for current SFQ fabrication, where the RSFQ logic integration density is only ${\sim}10^4$ gates/cm$^2$~\cite{razmkhah2024high}, the arithmetic circuits further stress-test the scalability of our approach. We use \textbf{qPlace}~\cite{7867080} as our baseline
for evaluating RSFQ placement quality. Our ablation studies examine linear versus quadratic objectives, level folding, timing-aware edge weighting, and sensitivity to $\lambda$. All methods are evaluated using qPALACE~\cite{10224066} with MIT-LL SFQ5ee technology \cite{tolpygo2014fabrication} and the gate library described in~\cite{razmkhah2025sce}. All placement algorithms were prototyped in Python using CPLEX for ILP and standard Python libraries for conjugate-gradient optimization, and were executed on a dual-socket AMD EPYC 7763 server with 128 cores and 2\,TB RAM.
For level folding, we set $\texttt{ITER\_MAX}=3$ and $\epsilon=0.1$. On BU32, the Python prototype takes 24 min, including 4.5 s for ILP and 70 s for QP. Most runtime is spent on Python overhead, including graph copying and traversal.

\subsection{Performance Comparison} 

As shown in Table~\ref{tab:qlevel_vs_qplace}, qLevelPlace improves CIL, area, and frequency while generally reducing HPWL.
More importantly, it improves the maximum clock frequency by 2.65$\times$ on average across the circuits completed by both methods, while producing routable layouts for all evaluated circuits. Moreover, note that qPlace does not yield routable layouts for c7552 and c5315 under its uniform row-spacing algorithm. In contrast,
our congestion-aware row spacing (Sec.~\ref{sec:splitter_placement}) explicitly reserves inter-row routing resources, which 
leads to successful routing of all tested designs.

\begin{figure}[t]
    \centering
    \subfloat[Effect of quadratic objective, lower is better.\label{fig:ablation-a}]{%
        \includegraphics[width=0.85\linewidth]{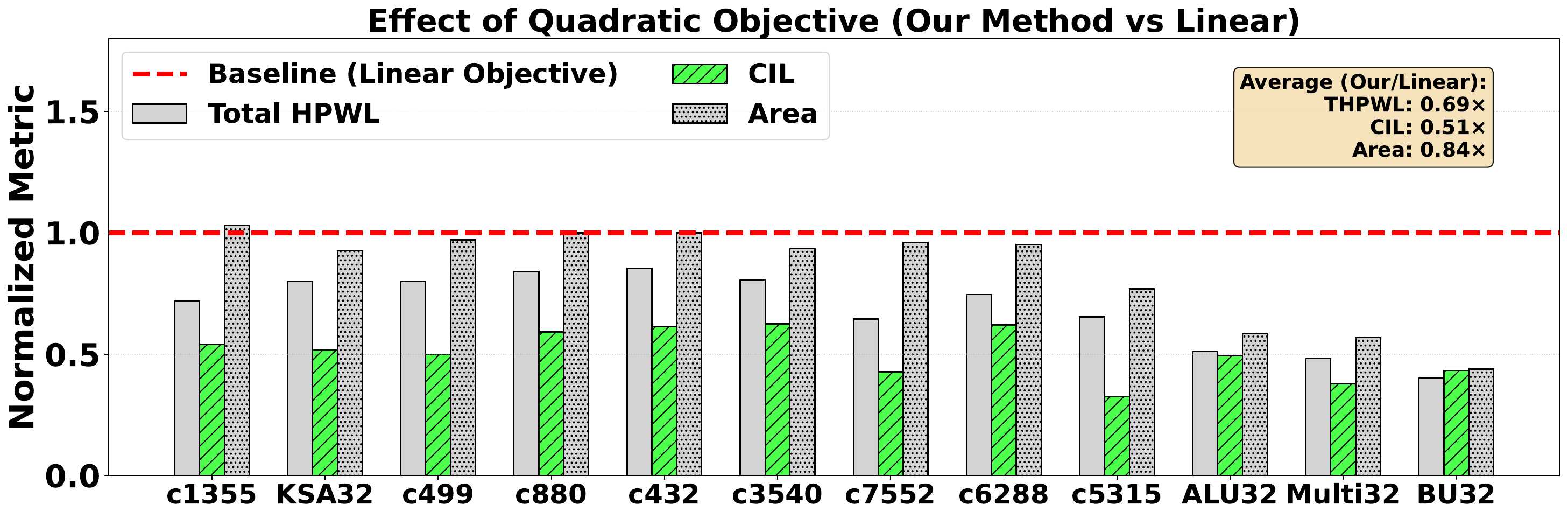}%
    }\\
    \subfloat[Effect of level folding, lower is better.\label{fig:ablation-b}]{%
        \includegraphics[width=0.85\linewidth]{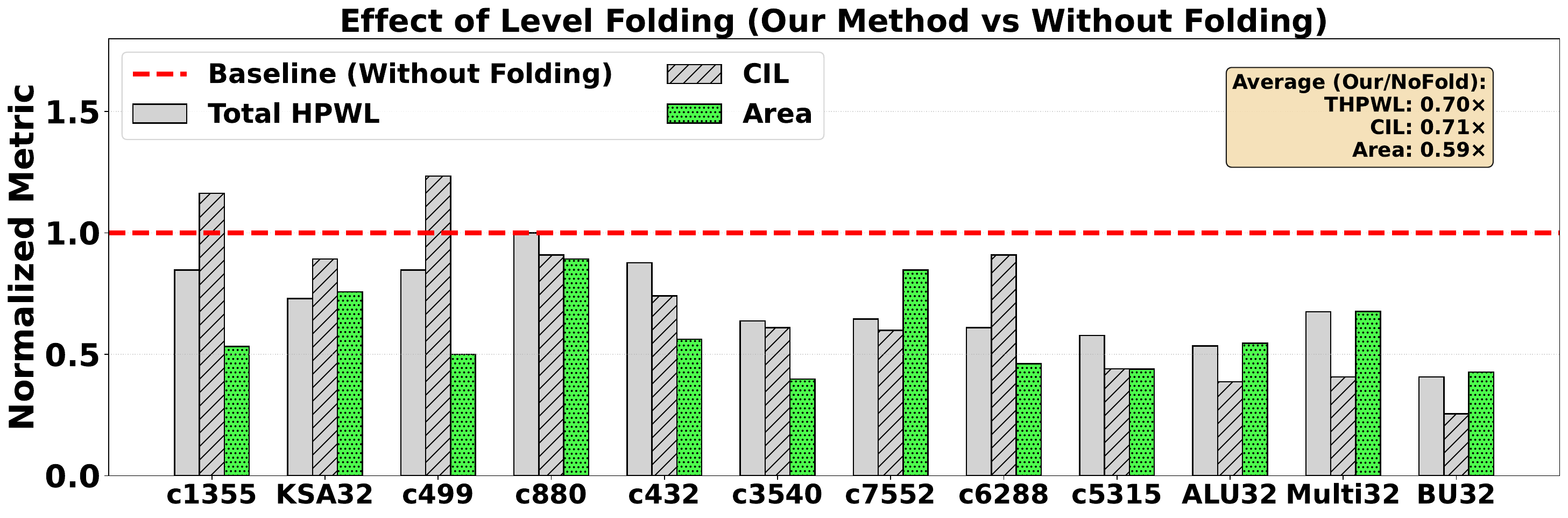}%
    }
    \\
    \subfloat[Sensitivity of anchor weight $\lambda$.\label{fig:ablation-c}]{%
    \includegraphics[width=0.80\linewidth]{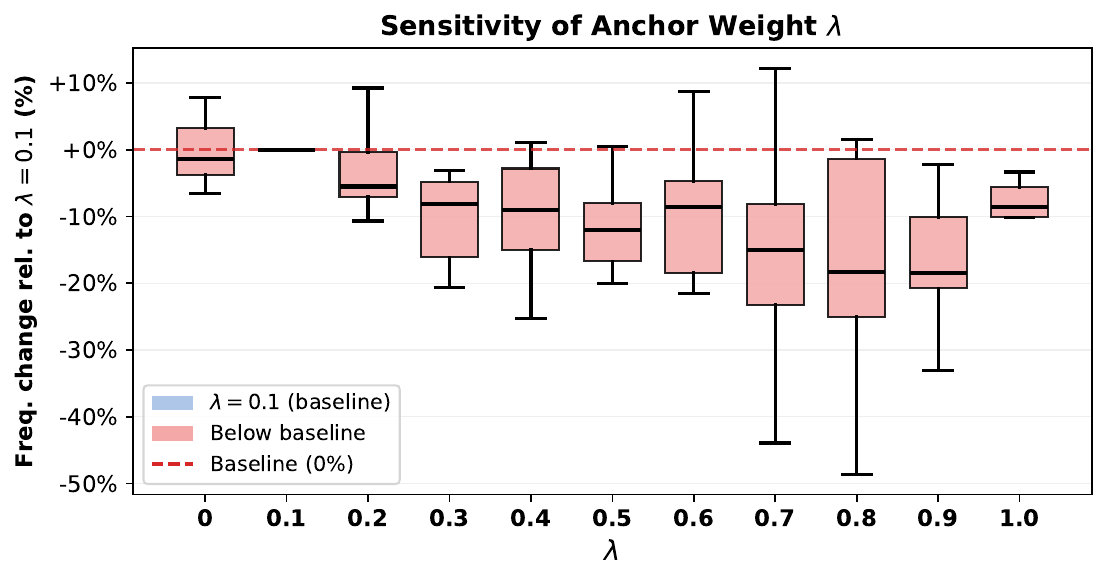}%
    }
    \vspace{-1em}
    \caption{Ablation study normalized to full qLevelPlace.}
    \Description{Three plots showing the effect of (a) linear optimization, (b) level folding, and (c) edge weighting on overall placement quality.}
    \label{fig:ablation}
    \vspace{-1em}
\end{figure}

\subsection{Ablation Studies}

Fig.~\ref{fig:ablation-a} isolates the contribution of the quadratic objective, whereas Fig.~\ref{fig:ablation-b} examines the role of level folding, and Fig.~\ref{fig:ablation-c} analyzes the sensitivity of frequency to $\lambda$. The quadratic objective reduces the estimated CIL by up to 3.06$\times$, as observed on c5315, confirming its effectiveness for RSFQ placement optimization. In fact, the linear model performs worse than the quadratic approach in both estimated CIL and total HPWL, which may seem counterintuitive, as linear objectives typically optimize total wire length. We attribute this to two factors. First, the linear model results in longer wire segments in the X-direction, increasing routing congestion and forcing our congestion-aware Y-axis spacing algorithm to expand the row spacing. On average, this Y-direction expansion accounts for 11.6\% of the per-benchmark increase in HPWL. Second, the row-by-row optimization with an approximated L$_1$ objective is more prone to converging to suboptimal local minima.

As shown in Fig.~\ref{fig:ablation-b}, disabling folding increases the layout areas of c3540 and BU32 to 2.51$\times$ and 2.34$\times$ those of the folded configurations, respectively. For BU32, it also increases total HPWL and CIL by 2.46$\times$ and 3.92$\times$, respectively. For smaller benchmarks, the layout area increases to 1.12--2.28$\times$ that of the folded configuration, while CIL may increase slightly because folding introduces additional rows and vertical separation. Nevertheless, CIL remains substantially lower than that of qPlace, and the benefits of folding become more pronounced for larger and more congested designs.

We further find the frequency to be insensitive to the anchor weight $\lambda$ (Eq.~\ref{eq:in_row_obj}) within $[0, 0.2]$ (Fig.~\ref{fig:ablation-c}), and adopt $\lambda{=}0.1$ throughout.

Finally, Table~\ref{tab:timing_weight} evaluates the effect of
timing-aware edge weighting.
For the critical stages identified from the without-weighting placement, timing-aware weighting reduces the estimated stage delay by 9.3\% on average and up to 17.7\% for BU32, the largest design, while reducing average CIL by 17\% and changing total HPWL only modestly.
The method selectively shortens delay-critical connections rather than merely minimizing global wirelength.

\section{Conclusions} 

Rather than using a uniform CMOS-style cost for all cells, qLevelPlace recognizes that clocked logic gates and unclocked splitters require distinct placement objectives: logic gates lie on independent timing paths and are placed to suppress worst-case interconnect length, whereas splitters act as constrained Steiner-like fanout points and follow the placed gates.
Decoupling the two lets placement focus on timing-critical gate-to-gate connections and, with a flexible level-to-row assignment, improves both timing and density.

On the benchmark circuits, qLevelPlace improves clock frequency by 2.65$\times$ over qPlace on average, showing that placement benefits substantially from reflecting the timing and fanout characteristics of RSFQ technology. The framework also extends naturally to sequential logic and to co-optimizing clocking, placement, and routing for related gate-level-clocked technologies such as AQFP.

\clearpage
\bibliographystyle{ACM-Reference-Format}
\bibliography{references}

\end{document}